# Wavefront shaping in complex media at 350 KHz with a 1D-to-2D transform

Omer Tzang[*1], Eyal Niv[*1], Sakshi Singh[1], Simon Labouesse[1], Greg Myatt[2], Rafael Piestun[1]

[1]*Department of Electrical, Computer, and Energy Engineering, University of Colorado Boulder, Colorado 80309, USA*

[2]*Silicon light machines, Sunnyvale, California, 94068, USA*

*[omer.tzang@colorado.edu](*omer.tzang@colorado.edu)

## Abstract

Controlling the propagation and interaction of light in complex media has sparked major interest in the last few years. Unfortunately, spatial light modulation devices suffer from limited speed that precludes real-time applications such as imaging in live tissue. To address this critical problem we introduce a phase-control technique to characterize complex media based on the use of fast 1D spatial light modulators and a 1D-to-2D transformation performed by the same medium being analyzed. We implement the concept using a micro-electro-mechanical grating light valve (GLV) with 1088 degrees of freedom modulated at 350 KHz, enabling unprecedented high-speed wavefront measurements. We continuously measure the transmission matrix, calculate the optimal wavefront and project a focus through various dynamic scattering samples in real-time, all within 2.4 ms per cycle. These results improve prior wavefront shaping modulation speed by more than an order of magnitude and open new opportunities for optical processing using 1D-to-2D transformations.

## Introduction

Recent developments in the field of wave-front shaping (WFS) have demonstrated control and optical focusing through complex media [1,2]. Coherent light in such media generates randomly scattered light fields that are seen as random 3D interference patterns, known as speckles [3]. Speckle fields can be manipulated by controlling the incident wave-front to generate enhanced intensity speckles at desired locations. Methods for focusing light through scattering media require an adaptive feedback process or phase conjugation to approximate the optical modes in the random media. Recent methods include wave-front optimization [4–6] and direct inversion of the measured transmission matrix [7].

Changes over time in random scattering media lead to speckle field changes. The speckle decorrelation time is defined as the time during which the correlation between the speckle field and an initial speckle field remains above a predetermined value. When focusing through scattering media using WFS, the speckle decorrelation reduces the intensity of the obtained focus over time. Dynamic biological tissues are extremely challenging for WFS focusing because blood flow reduces decorrelation times to the millisecond range. Typically, WFS is performed using high resolution liquid crystal (LC) spatial light modulators (SLM) and deformable mirrors. LC-SLMs devices are characterized by refresh rates in the order of 2-100ms. State of the art methodologies for faster wavefront optimization include micro-electro-mechanical system (MEMs) based mirror arrays [8], the use of binary deformable mirror devices (DMD) in phase modulation configuration [9,10], optical phase conjugation [11], and binary ferro-electric LC SLM [12]. These recent methodologies improve the focusing speed over traditional SLM based techniques but are still limited by the SLMs update speed and the use of binary phase wavefront that results in a lower enhancement potential. Moreover, phase conjugation focusing suffers from low SNR compared with feedback based WFS because it requires a light source inside or behind the scattering layer [13].

\* These authors contributed equally to this work

In this work, we investigate the use of fast one-dimensional (1D) SLMs for 2D WFS by taking advantage of the scattering medium to perform the 1D-to-2D transformation. In effect, a highly complex medium randomizes the degrees of freedom, by spreading individual 1D pixels or modes into uncorrelated 2D speckle fields, and hence provides a means to transform a 1D optical signal into a 2D field as graphically depicted in Fig. 1. We utilize a grating light valve (GLV), that is a high-speed 1D-SLM, to speed up feedback-based focusing through complex media. The GLV is a MEMs device, composed of thousands of free-standing silicon-nitride ribbons on a silicon chip, segmented into 1088 pixels, each composed of 6 ribbons as depicted in the top left inset of Fig.1. By electronically controlling the deflection of the ribbons, the GLV functions as a programmable 1D phase modulator. The GLV allows fast (<300ns) switching time, and high repetition rate operation (350 KHz in our case) along with continuous phase modulation. These properties allow three to four orders of magnitude faster operation compared to LC-SLMs, and more than one order of magnitude faster operation compared to binary amplitude DMDs and other binary phase modulators [8–12]. However, utilizing the GLV device for focusing in complex media requires tailored optical design as well as custom software – hardware implementation and signal processing.

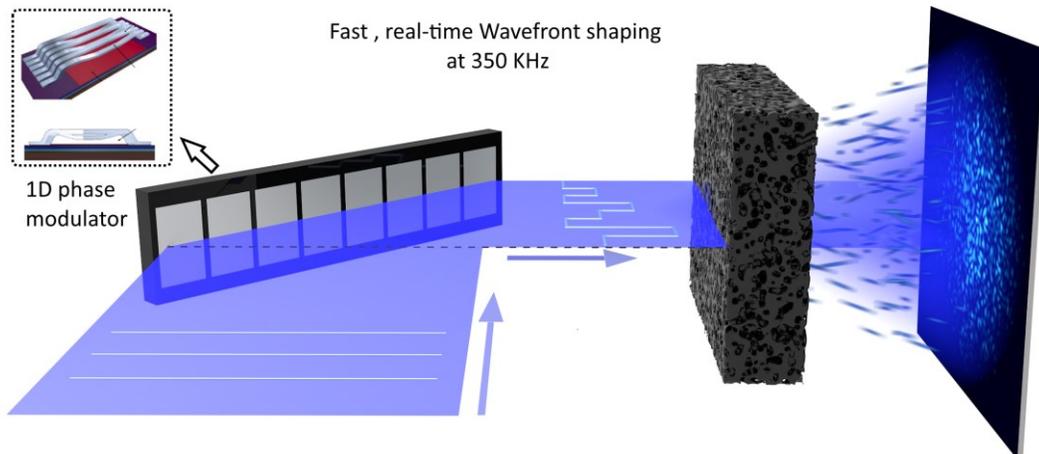

Figure 1: **Principle of 1D to 2D transformation for wavefront shaping with a 1D SLM**. A collimated and coherent laser beam illuminates the 1D SLM along a line - in our case a 1D MEMS phase modulator operating at 350 KHz, depicted in the top left inset. Using a tailored hardware and software implementation we measure the transmission matrix of complex media and focus light though it within 2.4 ms. The system operates continually in real time and allows examination of rapidly changing media.

This paper is organized as follows: We first describe and analyze the physical principle of 1D-2D transformation by a complex medium. We then describe the optical system utilized for WFS focusing in scattering media, and present experiments showing real-time fast focusing in static and dynamic media one order of magnitude faster than what was previously possible. We further demonstrate the technique is appropriate for focusing in multimode fibers (MMFs) and discuss advantages and limitations.

## 1D-to-2D transform via complex media

In this section, we investigate the basis for the use of a 1D phase modulator to enable 2D control of light propagating through complex media. The key observation is that the scattering medium performs a 1D-to-2D optical transformation by randomly distributing each 1D illuminating-mode into a 2D speckle field. Hence, assuming the speckle modes are fully developed, in the sense that the fields are random and uncorrelated in phase and amplitude, a 1D SLM provides the same 2D degrees of freedom as a 2D SLM with the same number of pixels. Mathematically, upon propagation through a thick random scatterer, the 1D and 2D wavefront modulators are essentially equivalent. Notwithstanding, it is important to analyze the physical effect of a relatively thin (or relatively weak) scatterer when the illumination shape is anisotropic, in our case, a 1D modulated light-line.

To model the far-field speckle shape generated by a scatterer of variable thickness with modulated light-line illumination, we represent the scattering medium with a TM correlation formulation [14] expanded to 2D fields (rather than 1-D as in [14]). The presence of speckle correlations, or Memory effect, manifests as a diffused elongated pattern at the near-field output of the scatterer and an elongated speckle grain, in the orthogonal direction, at the far-field. As the memory effect decreases with a thicker scatterer, the statistics of the speckle field becomes isotropic. A similar effect and behavior would be observed with a weak scatterer of given thickness as the scattering mean free path decreases. The TM model for anisotropic illumination takes into account random scattering by modelling an uncorrelated random matrix between the 2D input and output fields, $S_{2D}$. The memory effect is modeled by multiplying point by point $S_{2D}$ by a bound diagonal matrix, $G_{2D}$, that establishes different degrees of the memory effect. To generate $G_{2D}$, we use a Gaussian filter whose width, $\sigma$, is proportional to the degree of memory effect and in turn, the scatterer thickness [14]. Three examples of such TMs with different $\sigma$ values are depicted in Fig.2 (b,c,d). Note that the off-diagonal width increases with increasing $\sigma$. A 1D Hadamard phase pattern as the input field, shown in Fig.2(a), generates three different output fields, as depicted in Fig. 2 (f,g,h). The speckles in this far-field observation plane appear elongated and become statistically more isotropic as the scatterer width is increased.

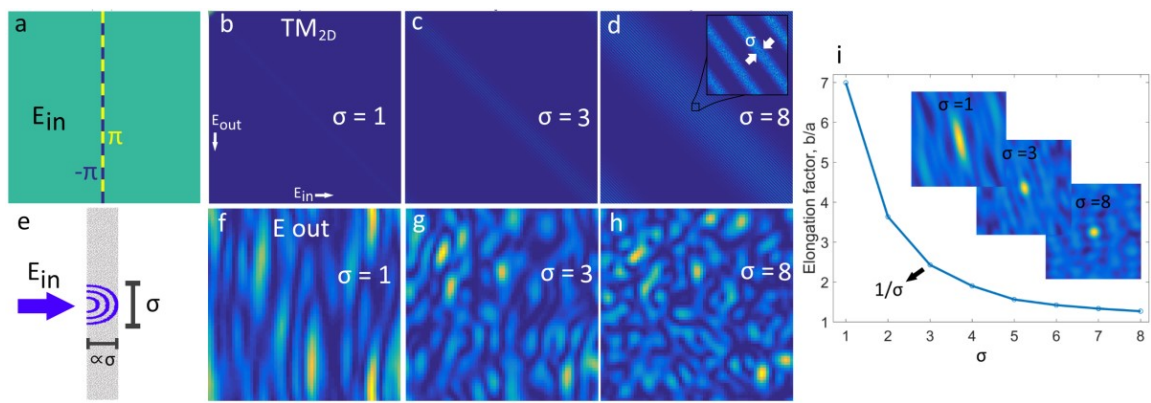

Figure 2: **Speckle shape for different scatterer widths upon line illumination**. (a) Line illumination input field. A line of alternating phases $(\pi, -\pi)$ with constant amplitude is depicted. (b,c,d) 2D transmission matrix corresponding to increasingly thicker scatterer, $\sigma = 1, 3, 8$, respectively. The images represent the absolute value of the transmission matrix and show a specific bound diagonal structure (e) A pencil beam illuminating a thin slab will cause a diffuse spot at the output surface, whose angular spread, $\sigma$ is of the order of the slab thickness. This angular spread along the orientation of the line illumination causes an elongation factor of $1/\sigma$ in the far field speckle. (f, g, h)**:** Simulation of speckle fields for the TMs shown in b,c,d respectively. The images represent the calculated intensity of the Fourier transform of the output near field. (i) Speckle elongation as a function of the scattering properties of the sample. At each point of the plot, we calculate the speckle elongation generated by the line illumination, as a function of $\sigma$, the width of the 2D Gaussian function used to model the TM. This Gaussian function directly relates to the scattering sample thickness, the Memory effect (angular correlations) and to the far-field shape of the speckles. We plot the elongation factor (ratio of major and minor axes) of the output field autocorrelation, averaged over 100 random realizations. The insets show characteristic shapes of speckle autocorrelation with different $\sigma$.

To quantify the speckle grain elongation, we calculate the autocorrelation of the speckle images and their corresponding average speckle grain size. We define the elongation factor as the ratio of the average grain's major and minor axes. Fig.2(i) shows the elongation factor drops as $1/\sigma$.

This is in agreement with the expected elongation in the far-field for a corresponding angular spread of σ in the near field image. Therefore, with anisotropic illumination of the scattering sample, the memory effect of the scatterer results in speckle elongation that varies according to the thickness or memory effect of the scattering medium.

## Experimental setup

The experimental setup is depicted in Fig.3. For illumination we use a 20mW, 460nm CW laser (New Focus, Vortex plus TLB 6800). The expanded beam generates a line illumination on the GLV (x-direction) after crossing a cylindrical lens. The GLV (F1088-P-HS) is placed at a reflection angle of ~10 degrees. A collimating cylindrical lens and a 6x demagnifying 4f system image the GLV (expanded in the y-direction) on the back focal plane of a 10x objective. The scattering sample is located at the focal plane of the objective, thus being illuminated with the 1D Fourier transform of the GLV phase distribution.

A 20x (NA = 0.4) objective images a plane behind the scattering sample. The speckle field propagates onto a pinhole placed before an avalanche photodiode (APD). The back objective and the pinhole size are selected to match the pinhole to the scattered speckle size. The APD voltage is digitized by a fast data acquisition card (DAQ) (Alazartech ,ATS9350), and sent to the computer where it is used to calculate the wavefront by a C++ program that controls all system computation and synchronization. A non-polarizing beam splitter reflects 1% of the light on a Camera (Point grey, Chameleon) to image the speckle field and focus spot.

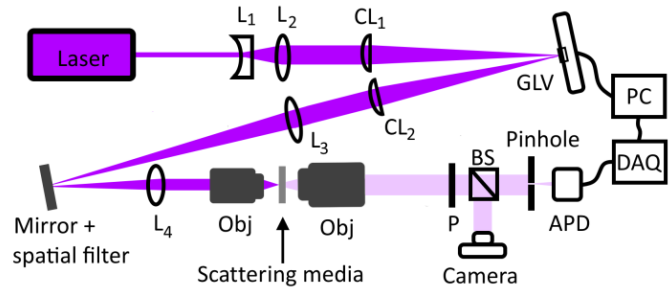

Figure 3: **Experimental setup**. A collimated 460nm laser beam illuminates the GLV. The imaging system of the GLV comprises a conjugate lens pair with an aperture in the Fourier plane between the lenses. The GLV image is created at the back aperture of the 10X objective and focused onto the scattering sample. A 20X objective images a plane behind the scattering sample. This image propagates to a pinhole placed before an APD. The APD signal is sent to a fast DAQ before being processed by a PC to create a focusing wavefront. A camera and beam splitter capture the focal spot image. GLV: grating light valve. L1-L4, are lenses with corresponding focal lengths of -50mm,300mm, 300mm and 50mm, CL1-2 are cylindrical lenses with focal lengths of 150mm. Obj –objectives. P: polarizer. BS: beamsplitter. APD: avalanche photodiode. DAQ: data acquisition card. PC: computer.

For WFS focusing, we select the transmission matrix method [9], because it uses a set of predefined phase masks that can be loaded to the GLV memory before operation. Using a preloaded set of phase masks minimizes the data transfer time between the GLV and the computer, allowing the GLV to display all preloaded images at its maximum frame rate. We characterize one column of the transmission matrix using three measurements per input mode, and calculate one focus spot per cycle. A key element enabling a high-speed system is the elimination of any computational or bandwidth bottlenecks in the feedback loop. We use high-bandwidth data transfer hardware, a dual-port data acquisition scheme, and a multi-threaded C++ application to speed up the focusing process. Fig.4(a) shows the intensity of a feedback speckle during WFS including: several TM measurements, high-speed data transfer and computation, and displaying the calculated phase mask on the GLV for focusing. Fig.4(b) shows the high-speed TM measurements.

The preloaded N input modes, where N= 256 or 512, are an orthogonal basis of phase patterns displayed on the center part of the GLV. We dedicate groups of GLV pixels for a modulated reference beam, displayed on both outer sides of the GLV, as shown in Fig.4(c). In the experiment, each mode interferes with three phase references (0, π/2, and π), displayed on the frame of the GLV and detected after propagation through the scattering medium by a fast detector. For precise phase measurements, we calibrate the GLV's voltage to phase transformation. After measuring the complex field response for all the input modes, the phase conjugated field is calculated and displayed on the GLV for focusing, similarly to what is done for 2D SLMs [15].

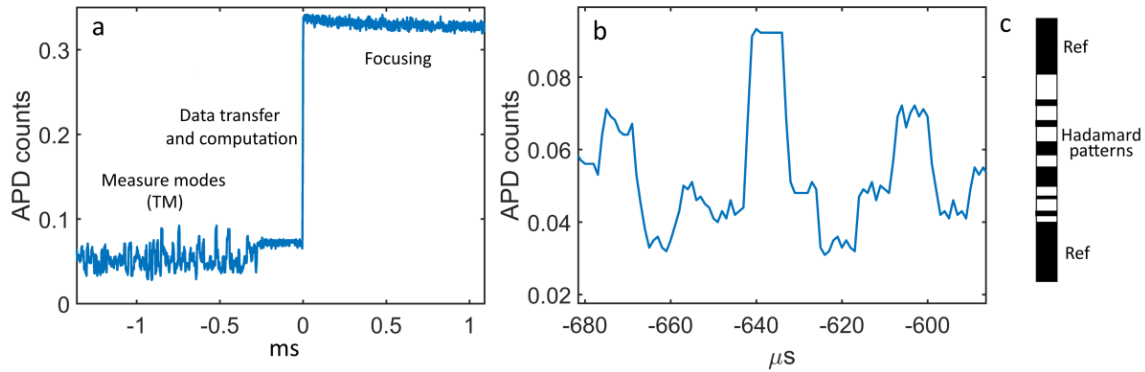

Figure 4: **High-speed WFS time signals**. (a) Timing of the system: Measured transmission matrix (TM) modes at 350Khz. Transfer data to computer, compute new mask, transfer data to GLV and project: 200 μs. Display optimized binary amplitude mask: 300 nanoseconds. For 256 modes the TM measurement time is 2.15ms and the real-time focusing cycle is 2.4ms, as depicted in Fig.5. (b) Zoom in of the measured modes (TM), shown in (a). The digitized APD signal show the intensity of few Hadamard basis modes, each of them interfered with three phase references. (c) An example of a phase distribution, for a single 1D Hadamard basis element surrounded by a phase reference for transmission matrix measurement.

## Results

Using a ground glass diffuser (Thorlabs, DG05-1500) as the scattering medium, we tested the GLV focusing system. Each phase mask was displayed for ~2.8 μs on the GLV that operated at 350 KHz. Thus, for N = 256, all 768 measurements for transmission matrix determination occurred in 2.15 ms. The APD signal is digitized and sent to the computer where the average intensity value for each measurement [16] is used to calculate the transmission matrix of the system, and display the focusing phase, all within an additional 150μs. Using the camera image, we calculated the enhancement as the peak intensity to average background ratio. Fig 5(a) illustrates the real-time focusing system using 256 modes. The focusing sequence takes 2.4ms and the focus is kept for another 5ms before the next measurement sequence. Fig 5(b) shows an example of a focus spot with 256 modes, demonstrating enhancement of x36 over the background level. Fig. 5.(c-d) shows the results obtained with 512 modes and signal enhancement of x60.

To test our system on controlled dynamic samples, we prepared scattering solutions with controlled viscosity and therefore varying speckle decorrelation times that mimic dynamic biological tissues. We also tested the system with various volume scattering samples including chicken breast, egg shell, and titanium-oxide nanoparticles, dried on a glass slide. All of these materials showed focusing enhancements in the same order as the glass diffuser.

Lastly, we tested our system for focusing of coherent light at the output of a MMF, a similar scenario to the speckle focusing in random scattering media. In MMFs, propagation of light is described by superpositions of propagating modes. Phase-velocity mode dispersion and random mode coupling arising from imperfections and bends contribute to creating complex 3D interference patterns observed as speckles at the fiber output. In the setup, we replace the scattering medium with a 30cm multimode fiber, including input and output coupling optics. The algorithm and system used for the MMF optimization were similar. For optimal results, we adjusted the coupling optics and imaged the GLV into the fiber with size-matched magnification that couple well all the GLV pixels. Fig. 5(e) shows a far field image of the fiber output during GLV optimization where a selected speckle is enhanced. The real time, high-speed control, is critical for maintaining a focus at the output of a rapidly moving fiber [10], for in-vivo imaging [17], or for controlling nonlinearities in MMFs [18].

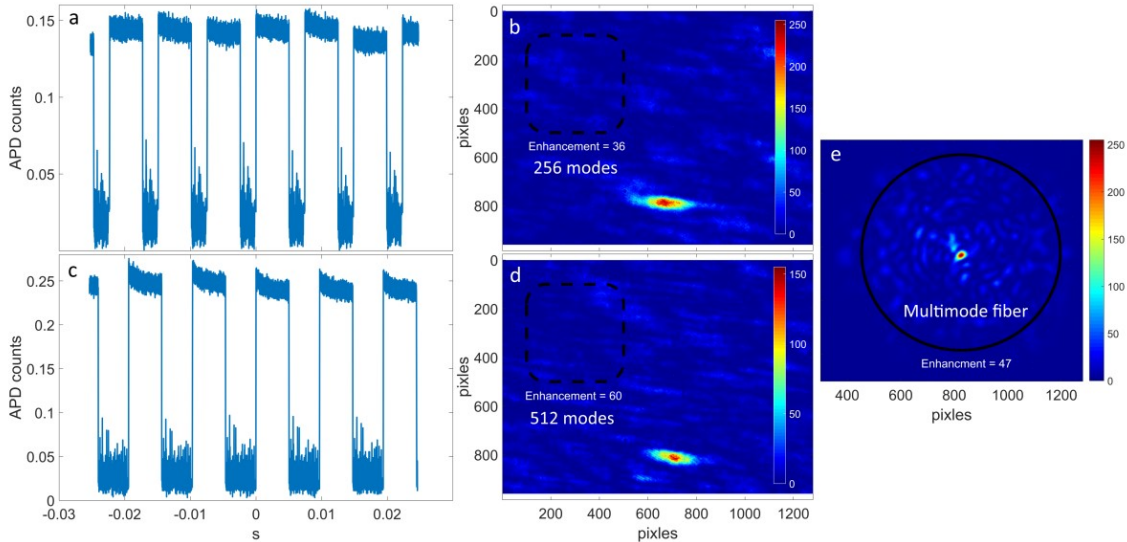

Figure 5: **System performance in terms of focus enhancement versus time**: (a) Real-time focusing using 256 modes. (b) Focal spot snap shot after wavefront optimization with 256 modes. The wave front was displayed continually and the reference beam blocked by displaying a high spatial frequency grating on the reference pixels and blocking its diffraction at the Fourier plane using a slit. (c) 512 modes. (d) Focal spot with 512 modes. (e) Focusing at the output of a graded-index MMF with diameter of 50µm, approximately 800 modes, and length of 30cm. Far-field image of the fiber output are recoded after wavefront optimization using 256 modes, at full GLV speed. The enhancements are comparable to the case of scattering media. The black circle indicates the fiber core.

Note that the speckles after the thin diffuser appear elongated as seen in Fig.5 (b,d) while at the fiber output, Fig.5(e), the speckles appear isotropic as expected from the random-media 1D-to-2D transform theory explained above.

## Discussion

The results presented herein show that proper system design enables the use of high-speed 1D SLMs for WFS. The concept of 1D-2D scattering transformation provides a framework for understanding and designing these WFS systems. Further system performance improvements are possible by taking into account the non-ideal characteristics of the GLV. In what follows, we discuss some of these areas open for improvement.

As expected, the intensity enhancement does not scale with N as predicted by theory in the ideal case [19] but still it is lower than the enhancement obtained using a phase-only LC-SLM and DMD for the same number of pixels [13]. Apart from sources of noise in the measurement [20], mechanical instabilities, and the common non-ideal aspects of SLMs such as distortion of the wavefront due to the pixel array structure, the existence of unmodulated light, and other wavefront distortions [6], the GLV incorporates a combination of additional features that require careful modelling when operated as a phase modulator. These include: low fill factor due to inter-pixel gaps, non-uniform bending of the MEMS ribbons along each pixel, settling time of the pixels at high speed operation, and reflections from the back surface of the device. Each GLV pixel contains 6 ribbons that move up and down together. The spacing between ribbons creates reflections from the back surface that interfere coherently and generate an additional higher frequency grating. The efficiency of this residual diffraction grating changes with ribbons' displacement and reaches 5-10% of a GLV pixel grating diffraction. Consequently, we observe that increasing the ratio of signal pixels that are static in the experiment vs reference pixels improves the focusing enhancement. For example, signal to reference pixel ratios of 70% and 95%, for the 256 and 512 modes, respectively, increased the overall SNR of mode interference, improved the accuracy of the measurement, and generated better enhancements. In addition, to improve the GLV performance, the back surface could be coated to minimize unwanted reflections.

Furthermore, the phase modulation range, determined by the GLV ribbon displacement, reached only ~$3/2\pi$ for our 460nm laser below the ideal $2\pi$, which matches 400nm at a reflection angle of 0 degrees. As a result, all phases between $3/2\pi$-$7/4\pi$ were set to $3/2\pi$ and all phases between $7/4\pi$ -$2\pi$ were set to $2\pi$. This limited phase range reduced the accuracy of the calculated wavefront. With improved GLV micro-mechanics, performance optimization, more accurate modeling of non-ideal behavior, and a wider GLV stroke, we expect the system to reach enhanced performance. In the optimization, we used either Hadamard or Fourier basis sets, with both showing similar performance. The illumination optics in our setup includes two cylindrical lenses that may cause astigmatism if their orientation is even slightly miss-matched and may contribute to the speckle elongation in thin samples. Additionally, the inhomogeneous Gaussian illumination of the 1D SLM using cylindrical lenses, even when the beam is expanded beyond the GLV, distributed a non-uniform intensity across the GLV pixels. When the TM mode are summed linearly in the focusing calculation, a phase error reduces the enhancement. This issue could be improved by flat illumination using a Powel lens or corrected computationally. Furthermore, the elongation of speckles could be eliminated even in very thin samples by using an additional diffuser, placed before the complex medium.

It is important to comment that the speed for focusing in scattering media is limited not only by the speed of the modulator and calculations but also by photon budget and SNR. We used a high transmission scattering layer and had sufficient, but not optimal, detector SNR for fast optimization of a single speckle. In other low SNR scenarios, such as weak fluorescence deep inside tissue, signal averaging could limit the overall speed.

## Conclusions

We have demonstrated high speed wavefront optimization for focusing through complex media using a fast 1D SLM with fast data acquisition and software adapted to the task. With this approach, we demonstrated an order of magnitude improvement in measurement speed over the current fastest feedback wavefront determination method and four orders of magnitude improvement over LC-SLM methods [2]. We also demonstrated real-time focusing through turbid materials during scan, focusing through dynamic scattering media, and controlling light at the output of MMFs. The improved speed is a significant technological step forward and holds potential for wide-field, video rate focusing and imaging in dynamic scattering media as well as high-speed control in MMFs. The concept of 1D-2D scattering transformation provides insight into the speckle correlations and shape in WFS with dimensionality mismatch, while guiding the design and utilization of WFS systems.

## Funding
This work was supported by NSF award 1548924 and 1611513 and NIH grant REY026436A.

## Acknowledgments
We thank Lars Eng, Alex Payne, and Yoshimi Hashimoto from Silicon Light Machines as well as Sylvain Gigan for useful discussions.